\documentclass[10pt,twocolumn]{IEEEtran}
\usepackage[T1]{fontenc}
\usepackage{amsfonts}
\usepackage{amssymb,amsmath,color,graphicx}
\usepackage{subfigure}
\usepackage{cite}
\usepackage{pdfpages}
\usepackage{comment}
\usepackage{verbatim}
\usepackage{algorithm}
\usepackage{algorithmic}
\usepackage{flushend}
\usepackage{placeins}
\usepackage{color,soul}
\usepackage{graphics} 
\usepackage{epsfig} 
\usepackage{times}
\usepackage{amsmath}
\usepackage{fix-cm}
\usepackage{bbold}
\usepackage[T1]{fontenc}

\usepackage{epstopdf}

\usepackage{amsmath,bm}

\usepackage{amssymb,amsmath,color,graphicx}
\usepackage{subfigure}
\usepackage{cite}
\usepackage{verbatim}
\usepackage{algorithm}
\usepackage{algorithmic}
\usepackage{setspace}
\usepackage{float}
\usepackage[T1]{fontenc}
\usepackage[english]{babel}
\usepackage{amssymb,amsmath,color,graphicx}
\usepackage{subfigure}
\usepackage{cite}
\usepackage{verbatim}
\usepackage{algorithm}
\usepackage{algorithmic}
\usepackage{setspace}
\usepackage{float}
\usepackage{dsfont}
\usepackage{bbold}
\usepackage{dsfont}
\usepackage{xr}
\usepackage{tabu}
\usepackage{tabularx}
\usepackage{longtable}
\usepackage{slashbox}
\usepackage[utf8]{inputenc}
\usepackage{pifont}
\usepackage{wasysym}
\usepackage{amssymb}
\newcommand{\specialcell}[2][c]{%
  \begin{tabular}[#1]{@{}c@{}}#2\end{tabular}}

\allowdisplaybreaks[1]
\usepackage{color}
\begin{document}

\title{\fontsize{20}{18} \selectfont \vspace{-4mm} A Decentralized  Trading Algorithm for an Electricity Market\\\vspace{1mm} with Generation  Uncertainty}
\author{\vspace{-2mm}\IEEEauthorblockN{Shahab Bahrami$^{\star}$ and M. Hadi Amini$^{\dagger}$}

\IEEEauthorblockA{$^{\star}$Department of Electrical and Computer Engineering, The University of British Columbia, Vancouver, Canada
\\$^{\dagger}$Department of Electrical and Computer Engineering, 
Carnegie Mellon University, Pittsburgh, PA, USA\\
 email:  bahramis@ece.ubc.ca, amini@cmu.edu, \vspace{-8mm}}}

\maketitle

\begin{abstract}

The uncertainties of the renewable generation units and the proliferation of price-responsive loads make it a challenge  for  independent system operators (ISOs) to manage the energy trading market in  the future power systems. A centralized energy market is not practical for the ISOs due to the high computational burden and violating the privacy of different entities, i.e., load aggregators and generators. In this paper, we propose a day-ahead \textit{decentralized} energy trading algorithm for a  grid with generation uncertainty.  To address the privacy issues, the ISO  determines some \textit{control signals} using the Lagrange relaxation technique   to motivate the entities towards  an  operating point that jointly optimize the   cost of load aggregators and profit of the generators, as well as the risk of the generation shortage of the renewable resources. More, specifically, we deploy the concept of conditional-value-at-risk (CVaR) to minimize the risk of renewable generation shortage.  The performance of the proposed algorithm is evaluated on an IEEE 30-bus test system. Results show that the proposed decentralized algorithm converges to the solution of the ISO's centralized problem in 45 iterations. It also benefits both the load aggregators by reducing their cost by 18$\%$ and the generators by increasing their profit by 17.1$\%$.

\vspace{2mm}
\end{abstract}

\vspace{-2mm}
\section{Introduction} \label{sec:introduction}

 One goal of the future  smart grid is to provide the necessary infrastructure towards integration of renewable  generators (e.g., wind turbine,  photovoltaic (PV) panel) as the environmentally friendly alternatives for the fossil fuel-based generators. Meanwhile,  the  infrastructure in smart grid facilitates the participation of demand side in the energy management programs such as demand response (DR)~\cite{kar14}.
 
The  \textit{optimal power flow} (OPF) analysis is widely used by independent system operators (ISOs)  to optimally  dispatch the available generation
portfolio to meet the load demand. There are challenges to address the OPF analysis in a grid with renewable energy generators and price-responsive load demands.  First, addressing the OPF problem by the ISO can violate the  entities' privacy in a competitive energy market, e.g., by revealing the demand information  of the load aggregators and cost of the generators. Second,  the OPF problem is nonlinear and nonconvex problem, which is difficult to be solved. Even by linearizing the problem (e.g., in DC OPF), it  can  be computationally difficult in large networks, especially when the number of decision variables  increases by participating the price-responsive load aggregators in the energy market. Third, the uncertainty of the renewable generation puts the generation-load balance at risk, especially when the ISO does not have access to the accurate historical generation data of the privately owned  renewable generators.

There have been some efforts in the power system operation literature to tackle the above-mentioned challenges. We divide the related works into three thread. The first thread is concerned with  privacy issues and focuses on designing a distributed energy planning programs for a system with multiple suppliers and multiple users using  the evolutionary game  \cite{Bo}, Stackelberg game \cite{maharjan}, dual decomposition method  \cite{deng}, supply bidding mechanism \cite{farhad}, and hierarchical bidding  \cite{ucdr4}. These approaches may not be easily implementable in practice, since they ignore  the physical constraints imposed by  the topology and operation of the power network. The second thread is concerned with including  the power flow constraints in the decentralized system planning procedure using different techniques such as the primal-dual algorithm  \cite{DDCOPF}, convex relaxation  \cite{opfdr1, opfdr3, new1}, quadratic programming \cite{new2}, alternating
direction method of multipliers (ADMM) \cite{new4, new6,new8}, and Lagrange relaxation method \cite{new5,new7}. In these works, however, the uncertainty issues related to integrating renewable generators remain a challenge. The third thread is concerned with 
 addressing the uncertainty issues by using different techniques such as chance constrained optimization \cite{online1,online2,online4}, successive constraint enforcement \cite{online3}, fuzzy systems \cite{online5}, stochastic optimization \cite{online6}, and risk management \cite{online7,pedramshahab}. The proposed models are  centralized, and  cannot address the  privacy  and computational complexity challenges.

In this paper, we focus on extending the works in the third thread by designing a distributed energy trading algorithm  in a day-ahead electricity market with renewable energy generators and active participation of the load aggregators in load management programs. The ISO, load aggregators, and generators use the smart grid's communication infrastructure  to execute the proposed distributed  algorithm and jointly optimize the generators' profit  and the load aggregators'  cost. The generators' and load aggregators  solves their own optimization problems using some locally available information. Hence, the privacy of each entity is protected.  The main challenge is to determine proper control signals among the ISO, generators, and load aggregators that enforce the proposed distributed  algorithm converges to the optimal solution of the  \textit{centralized} problem for an ISO with complete information.

 The main contributions of this paper are as follows:
\begin{itemize}
\item \textit{Risk Evaluation}: Inspired by the works in \cite{online7,pedramshahab},  we deploy a penalty based on the conditional value-at-risk (CVaR) into the objective functions of the ISO's centralized problem and the generators' local problems to address the uncertainties of the renewable generation.  It enables the generators with renewable plants to sell electricity and gain profit, while limiting the risk of high  generation shortage within a certain confidence level. The ISO can also reduce the risk of  generation-load mismatch. We discuss how to deal with the difference between the risk aversion levels of the ISO and generators.
\item \textit{Distributed Algorithm Design}: In order to protect the privacy of load aggregators and generators, and to address the computational complexity of the centralized approach, we propose a decentralized algorithm based on the Lagrange relaxation method \cite{new5,new7} that can be executed by the entities in the day-ahead market. A  load aggregator to manage the controllable loads of its customers. A generator determine its risk minimizing  generation level. Meanwhile, the ISO can meet the  power flow constraints in
the grid,  thereby achieving a triple-win result.
\item \textit{Performance Evaluation}: Simulation results on an IEEE 30-bus test system show that the proposed decentralized algorithm  can converge to the global optimal solution to the centralized problem of  the ISO in about $45$ iterations. The proposed algorithm also benefits both the load aggregators by reducing their cost by $18\%$ and the generators by increasing their profit by $17.1\%$ and reducing the peak-to-average ratio (PAR) by $15\%$. When compared with the centralized approach, our algorithm has a significantly  lower computational time. when compared with  a centralized method with AC power flow in different test systems, our approach has a lower running time at the cost of $3\%$ to $8\%$ error due to the DC power flow approximation.
\end{itemize}

 The rest of this paper is organized as follows. Section \ref{s2} introduces the models for load aggregators, generators and ISO. In Section \ref{s3} we propose a decentralized algorithm  to determine the  energy market equilibrium.  Section IV provides  simulation results. Section \ref{s6}  concludes the paper. {Appendices A and B can be found in the supplementary document.}

\vspace{-3mm}

\section{System Model}\label{s2}
Consider a \textit{day-ahead}  energy market with a set  $\mathcal{N}$ of load aggregators and a set $\mathcal{M}$ of  generators.
Each \textit{load aggregator} is responsible for managing the load demand on be half of its electricity users.  Each generator sells electricity to the market. The load aggregators and  generators use a two-way communication infrastructure  to exchange information with the ISO.  The load aggregators and generators are also connected to each other  within the transmission network. The ISO  monitors the power flow through the transmission lines over the trading horizon $\mathcal{H}=\{1,\dots,H\}$, where $H$ is the number of time slots with equal length, e.g.,  one hour.

  To avoid the potential of confusion, we assume that each bus  has either a load aggregator or generator. If a load aggregator  and a generator  are connected to the same bus, we divide that bus into two buses connected to each other through a line with zero impedance. If neither load aggregator nor generator is connected to a bus, we add a virtual load aggregator with zero demand to that bus. It enables us to denote the set of buses by $\mathcal{N}\cup \mathcal{M}$ and  refer a load aggregator or a generator by  its bus number. We use notation $\mathcal{L}\subseteq (\mathcal{N}\cup \mathcal{M})\times  (\mathcal{N}\cup \mathcal{M})$ to denote the set of  transmission lines. 
In the energy market, the ISO provides each entity with the day-ahead market price values for trading the electricity. Let $\rho_{b,h}$  denote the price in time slot $h\in\mathcal{H}$ for the entity connected to bus $b\in \mathcal{N}\cup \mathcal{M}$. Let vector $\boldsymbol{\rho}_{b}=(\rho_{b,h},\;h\in\mathcal{H})$ denote the day-ahead price profile.  In the  rest of this section, we will provide the models for the load aggregators, generators, and ISO.

\subsubsection{Load Aggregator's Model} 
Load aggregator $i\in\mathcal{N}$ is responsible for managing the load demand of its users. The load demand $l_{i,h}$ in bus $i$ in time slot $h$ consists of uncontrollable (or baseload) demand $l^{\text{b}}_{i,h}$ and controllable demand $l^{\text{c}}_{i,h}$.  The controllable demand comprises a set $\mathcal{A}_i$ of  different types of controllable loads for residential, commercial, and industrial users connected to bus $i$. Controllable load $a\in\mathcal{A}_i$ has a \text{scheduling horizon} $\mathcal{H}_{a,i}\subseteq \mathcal{H}$ that defines the time interval, in which  the load should be scheduled. We further divide the controllable loads into two types based on their  characteristics.


A controllable load  of \textit{type 1} have hard deadline. It should be operated  within the scheduling horizon and turned off  in other time slots. Examples  include the household's electric vehicle (EV) and production line of an industry. A controllable load of \textit{type 2}  is more flexible. it can be operated in  time slots out of the scheduling horizon at the cost of high discomfort for the user, e.g.,  lighting in households, packing process in industries, and air conditioner in commercial buildings. 
 Let $x_{a,i,h}$ denote the demand of controllable load $a\in \mathcal{A}_i$ in time slot $h$. Let $\mathcal{A}_i^1$ and $\mathcal{A}_i^2$ denote the set of controllable loads of type 1 and 2 for load aggeragtor $i\in\mathcal{N}$, respectively. We have  
\begin{subequations} 
  \begin{align}
 &\!\!\!\! x_{a,i,h}=0,\,\hspace{2.6cm} a\in\mathcal
  A_i^1,\, h\not\in\mathcal{H}_{a,i},\label{0}\\ 
  &\!\!\!\! x_{a,i,h}\ge 0,\,\hspace{2.6cm} a\in\mathcal
  A_i^2,\, h\not\in\mathcal{H}_{a,i},\label{01}\\ 
  &\!\!\!\!x_{a,i,h}^{\text{min}}\!\leq\! x_{a,i,h}\!\leq \!x_{a,i,h}^{\text{max}}, \hspace{1cm}  a\in\mathcal
  A_i^1\cup \mathcal{A}^2_i,\, h\in\mathcal{H}_{a,i},\label{1}\\
&\!\!\!\!X_{a,i,h}^{\text{min}}\leq \textstyle\sum_{h\in\mathcal{H}}x_{a,i,h}\leq X_{a,i}^{\text{max}},\hspace{0.4cm}  a\in\mathcal
  A_i^1\cup \mathcal{A}^2_i.\label{2}
\end{align}
\end{subequations}

 Scheduling the controllable loads results in a discomfort cost for the users. The discomfort cost  for  type 1 loads depends only on the total power consumption deviation from the desirable value (e.g., the total charging level of an EV is important).  For the scheduled power consumption profile $\bm{x}_{a,i}=(x_{a,i,h},\,h\in\mathcal{H})$ and desirable profile $\bm{x}^{\text{des}}_{a,i}=(x^{\text{des}}_{a,i,h},\,h\in\mathcal{H})$, the  discomfort cost $\Omega_{a,i}(\bm{x}_{a,i})=\omega_{a,i} \big(\sum_{h\in\mathcal{H}_{a,i}}(x_{a,i,h}-x^{\text{des}}_{a,i,h})\big)^2$ with  nonnegative constant $\omega_{a,i}$ is a viable candidate. The discomfort cost for type 2 loads depends on both the amount of power consumption  and the time of consuming the power.  The  discomfort cost  $\Omega_{a,i}(\bm{x}_{a,i})=\sum_{h\in\mathcal{H}_{a,i}}\omega_{a,i,h}(x_{a,i,h}-x^{\text{des}}_{a,i,h})^2+\sum_{h\not\in\mathcal{H}_{a,i}}\omega'_{a,i,h}x_{a,i,h}$ with   \textit{time dependent} nonnegative coefficients $\omega_{a,i,h}$ and $\omega'_{a,i,h}$, $\omega'_{a,i,h}\gg\omega_{a,i,h}$ is a viable candidate.

Let $\bm{x}_i=(\bm{x}_{a,i},\,a\in\mathcal{A}_i)$ denote the profile of power demands over all loads of load aggregator $i$. 
%
Load aggregator $i$ aims to minimize the total cost  $C^{\text{agg}}_i(\boldsymbol{x}_i)$, which includes the discomfort cost $\Omega_{i}(\bm{x}_{i})= \sum_{a\in\mathcal{A}_i}\Omega_{a,i}(\bm{x}_{a,i})$ and the payment to the ISO to meet uncontrollable demand $l^{\text{b}}_{i,h}$ and controllable demand $l^{\text{c}}_{i,h}=\sum_{a\in\mathcal{A}_i}x_{a,i,h}$. For $i\in\mathcal{N}$, we have
\begin{align}\label{costt}
C^{\text{agg}}_i(\boldsymbol{x}_i)=\Omega_{i}(\bm{x}_{i})+\textstyle\sum_{h\in\mathcal{H}} \rho_{i,h}\big(l^{\text{b}}_{i,h}+l^{\text{c}}_{i,h}\big).
\end{align}

Let $\mathcal{X}_i$ denote the feasible space defined by constraints (\ref{0})$-$(\ref{2}) for load aggregator $i$. Load aggregator $i\in\mathcal{N}$ solves the following optimization problem.
 \begin{subequations}\label{agg_prob} 
\begin{align}
	&\hspace{1cm}\underset{\boldsymbol{x}_i}{\text{minimize}}\;\; C^{\text{agg}}_i(\boldsymbol{x}_i) & 	\\
&\hspace{1cm}\text{subject to} \;\; \boldsymbol{x}_i\in\mathcal{X}_i. \hspace{0cm}\label{demand}
\end{align}
\end{subequations}

\subsubsection{Generator's Model} Generator $j\in\mathcal{M}$ sells  $p^{\text{conv}}_{j,h}$ of its conventional unit's output power to    in times slot $h$.   Let $\boldsymbol{p}^{\text{conv}}_j=(p^{\text{conv}}_{j,h},\,h\in\mathcal{H})$ denote the conventional unit's generation profile  for generator $j$. 
      The  cost  of a conventional unit in time slot $h$ is generally an increasing convex function of  $p^{\text{conv}}_{j,h}$ \cite{genfunc}.  The class of quadratic generation cost functions $C^{\text{conv}}_{j,h}(p^{\text{conv}}_{j,h})=a_{j2}(p^{\text{conv}}_{j,h})^2+a_{j1}p^{\text{conv}}_{j,h}+a_{j0}$ is well-known \cite{24}. 

A generator can also use renewable units, such as solar and wind to benefit from their zero generation cost. Due to the uncertainty, generator $j$  offers a renewable generation profile $\boldsymbol{p}^{\text{ren}}_j=(p^{\text{ren}}_{j,h},\,h\in\mathcal{H})$, such that the \textit{risk} of deviation from the \textit{actual} generation profile $\widehat{\boldsymbol{p}}^{\,\text{ren}}_j$  is minimized. Specifically,  to prevent the generators  from non-credible high generation offers in the   market, the ISO charges generator $j$ with renewable units by the price $\theta_{j,h},\,h\in\mathcal{H}$ (cents/kW), when its actual generation $\widehat{p}^{\,\text{ren}}_{j,h}$ is \textit{lower} than its offer $p^{\text{ren}}_{j,h}$. A generator with renewable units can use  risk measures such as the value-at-risk (VaR) and conditional value-at-risk (CVaR) to limit the risk of generation shortage. Let $\Delta_{j,h}(p_{j,h}, \widehat{p}_{j,h})$ denote a  function that captures the penalty for generation shortage in time slot $h$ for generator $j$ with renewable units\cite{rockafellar2000}.
It is defined~as
\begin{align}
&\Delta_{j,h}(p^{\text{ren}}_{j,h}, \widehat{p}^{\,\text{ren}}_{j,h}) =  \theta_{j,h} \left[p^{\text{ren}}_{j,h}- \widehat{p}^{\,\text{ren}}_{j,h}\right]^{+}, \hspace{1mm} h\in\mathcal{H},\; j\in\mathcal{M},\label{opt2}
\end{align}
where $[\cdot]^{+} = \text{max}\{\cdot,0\}$.   
$\Delta_{i,h}(p^{\text{ren}}_{j,h}, \widehat{p}^{\,\text{ren}}_{j,h})$ is a random variable, since the actual generation  $\widehat{p}^{\,\text{ren}}_{j,h}$ is a stochastic process. 
Under a given confidence level $\beta_j \in (0,1)$ and the offered generation level $p^{\text{ren}}_{j,h}$ in time slot $h$,  the VaR for generator $j$ 
is defined as the minimum threshold cost $\alpha_{j,h}$, for which the probability of generation shortage of generator $j$  being less than $\alpha_{j,h}$ is at least $\beta_j$ \cite{rockafellar2000}. 
That is,
\begin{equation}
\text{VaR}_{j,h}^{\beta_j}\left(p^{\text{ren}}_{j,h}\right) = \text{min} \left\{\alpha_{j,h}\;|\; \text{Pr}\left\{\Delta_{i,h}(\cdot)\leq \alpha_{j,h} \right\} \geq \beta_j \right\}.
\end{equation}

Due to the non-convexity, it is difficult to minimize the VaR. The CVaR is an alternative risk measure, which is convex and can be optimized using sampling techniques. The CVaR for generator $j$ with renewable units in time slot $h$ is defined as the expected value of the generation shortage cost  $\Delta_{j,h}(p^{\text{ren}}_{j,h}, \widehat{p}^{\,\text{ren}}_{j,h})$ when only the costs that are greater than or equal to $\text{VaR}_{j,h}^{\beta_j}(p^{\text{ren}}_{j,h})$ are considered \cite{rockafellar2000}. 
That is,
\begin{align}
\label{CVAR}
&\!\!\text{CVaR}_{j,h}^{\beta_j}\left(p^{\text{ren}}_{j,h}\right)=\text{E}\left\{\Delta_{j,h}(\cdot)\big| \Delta_{j,h}(\cdot) \geq \text{VaR}_{j,h}^{\beta_j}\left(p^{\text{ren}}_{j,h}\right)\right\}.
\end{align}
 It is possible to estimate the CVaR  by adopting sample average approximation (SAA) technique \cite{rockafellar2000}. 
Samples of the random variable $\widehat{p}^{\,\text{ren}}_{j,h}$ for generator $j$ with renewable units in time slot $h$ can be observed from the historical record.
Consider the set $\mathcal{K}= \{1,\dots,K\}$ of $K$ samples of the random variable $\widehat{p}^{\,\text{ren}}_{j,h}$. Let $p^{\text{ren},k}_{j,h}$ denote the $k^{\text{th}}$ sample of   $\widehat{p}^{\,\text{ren}}_{j,h}$ for generator  $j$ in time slot $h$. The CVaR  in (\ref{CVAR}) can be approximated by $ \text{CVaR}_{j,h}^{\beta_j}(p^{\text{ren}}_{j,h}) \approx \underset{\alpha_{j,h}}{\text{min}} ~ \widetilde{\mathcal{C}}_{j,h}^{\beta_j}(p^{\text{ren}}_{j,h}, \alpha_{j,h})$, where
\begin{equation}
\label{gamma2}
\!\widetilde{\mathcal{C}}_{j,h}^{\beta_j}\left(p^{\text{ren}}_{j,h}, \alpha_{j,h}\right)= \alpha_{j,h}\! +\! \displaystyle \sum_{k\in\mathcal{K}}  \!\frac{\big[ \Delta_{j,h}(p^{\text{ren}}_{j,h}, p^{\text{ren},k}_{j,h}) \!-\!\alpha_{j,h} \big]^{\!+}}{K(1 \!- \! \beta_j)}  \!~.
\end{equation}

Under the given offer $p^{\text{ren}}_{j,h}$, we can use the historical samples of the renewable unit's output power in each time slot to compute $\widetilde{\mathcal{C}}_{j,h}^{\beta_j}(p^{\text{ren}}_{j,h}, \alpha_{j,h})$.   For further simplification of the objective function (\ref{supprob}), we replace the terms $\big[ \Delta_{j,h}(p^{\text{ren}}_{j,h}, p^{\text{ren},k}_{j,h}) \!-\!\alpha_{j,h} \big]^{\!+}$ in  (\ref{gamma2}) with the auxiliary variables $\eta_{j,h}^k$ for every sample $k\in\mathcal{K}$ in  time slot $h\in\mathcal{H}$. We include $ \theta_{j,h}(p_{j,h}-p^{k}_{j,h})-\alpha_{j,h}\leq\eta_{j,h}^k$ into the constraint set. We define $\boldsymbol{\eta}_{j,h}=(\eta_{j,h}^k,\,k\in\mathcal{K})$.  Hence, (\ref{gamma2}) can be rewritten as 
 \begin{equation}
\label{gamma3}
\widetilde{\mathcal{C}}_{j,h}^{\beta_j}\left(\boldsymbol{\eta}_{j,h}, \alpha_{j,h}\right)= \alpha_{j,h} + \displaystyle {\sum_{k\in\mathcal{K}} } \dfrac{\eta^k_{j,h}}{K(1 \!- \! \beta_j)}  \! ~ .
\end{equation}

 The  objective of  a generator is to maximize its profit $\pi^{\text{gen}}_j$, which is    the revenue from selling electricity in the market with prices $\rho_{j,h},\,h\in\mathcal{H}$ minus the generation cost and the financial risk  CVaR in (\ref{gamma3}). We define decision vector ${\psi}_j=(\boldsymbol{p}^{\text{conv}}_j,\boldsymbol{p}^{\text{ren}}_j, (\boldsymbol{\eta}_{j,h}, \alpha_{j,h},h\in\mathcal{H}))$ for generator $j$. The generator's profit is obtained as follows.
\begin{align}\label{supprob}
&\pi^{\text{gen}}_j({\psi}_j)=\nonumber\\&\sum_{h\in\mathcal{H}}\!\big((p^{\text{conv}}_{j,h}+p^{\text{ren}}_{j,h})\rho_{j,h}\!-\!C^{\text{conv}}_{j,h}(p_{j,h})\!-\!\widetilde{\mathcal{C}}_{j,h}^{\beta_j}\!\left(\boldsymbol{\eta}_{j,h}, \alpha_{j,h}\right)\!\big).
\end{align}
 The local optimization problem for generator $j\in\mathcal{M}$ is 
 \begin{subequations}\label{util_problem}
 \begin{align}\label{util_obj}
&\displaystyle \underset{{\psi}_j} {\text{maximize}}\,\;\pi^{\text{gen}}_j({\psi}_j)\\
&\text{subject to}\;\;p^{\text{min}}_{j}\leq p^{\text{conv}}_{j,h}\leq p^{\text{max}}_{j} ,\hspace{1.8cm}  h\in\mathcal{H},\label{gen1}\\
    &\hspace{0cm} \theta_{j,h}(p^{\text{ren}}_{j,h}-p^{\text{ren},k}_{j,h})-\alpha_{j,h}\leq\eta_{j,h}^k,\; \;\;\;\;k\in\mathcal{K}, \,h\in\mathcal{H},\label{gen3}\\
  &\hspace{2.0cm}0\leq \alpha_{j,h},\eta_{j,h},\hspace{2.25cm}  h\in\mathcal{H}.\label{gen2}
\end{align}
\end{subequations}

     
\subsubsection{ISO's Model}




A variety of different objectives can be considered for the ISO in the energy market. In this paper, we consider the objective of minimizing the social cost with the risk of generation shortage of all renewable generators over the planning horizon. The social cost is the generation cost of the \textit{conventional} generators  plus the discomfort cost of load aggregators.
Regarding the risk minimization, the ISO can consider the CVaR  similar to (\ref{opt2})$-$(\ref{gamma3}). Remind that from (\ref{opt2}),  the penalty for the  generation shortage is important for a generator. However, from the ISO's perspective, the value of generation shortage of the renewable unit is important. That is, from the ISO's  point of view, the cost of generation shortage in time slot $h$ corresponding to generator $j$  is $\Delta^{\text{ISO}}_{j,h}(p^{\text{ren}}_{j,h}, \widehat{p}^{\,\text{ren}}_{j,h}) =   [p^{\text{ren}}_{j,h}- \widehat{p}^{\,\text{ren}}_{j,h}]^{+}$.  Moreover, the ISO may consider a confidence level $\beta_j^{\text{ISO}}$ different from the confidence level $\beta_j$  of generator~$j$. 

Similar to (\ref{gamma3}), we can formulate  $\widetilde{\mathcal{C}}_{j,h}^{\,\beta^{\text{ISO}}_j}(\boldsymbol{\eta}^{\text{ISO}}_{j,h}, \alpha^{\text{ISO}}_{j,h})$ as the approximate CVaR function that the ISO assigns to generator $j$. Here, $\eta^{\text{ISO},k}_{j,h}$ is the auxiliary variable corresponding to  term $ [p^{\text{ren}}_{j,h}- \widehat{p}^{\,\text{ren}}_{j,h}]^{+}$ and  $\boldsymbol{\eta}^{\text{ISO}}_{j,h}=(\eta^{\text{ISO},k}_{j,h},\,k\in\mathcal{K})$ is the vector of auxiliary variables for samples $k\in\mathcal{K}$ of the historical data for generator $j$ in time slot $h$. We have
 \begin{equation}
\label{gamma4}
\widetilde{\mathcal{C}}_{j,h}^{\beta^{\text{ISO}}_j}\left(\boldsymbol{\eta}^{\text{ISO}}_{j,h}, \alpha^{\text{ISO}}_{j,h}\right)= \alpha^{\text{ISO}}_{j,h} + \displaystyle \sum_{k\in\mathcal{K}}  \frac{\eta^{\text{ISO},k}_{j,h}}{K(1 \!- \! \beta^{\text{ISO}}_j)}.
\end{equation}

As a netural entity,  the ISO  uses the DC power flow  model as an acceptable framework to determine the active power flow through the transmission lines and the generation-load balance in  transmission systems~\cite{dcopf04, dcpf09}. Let  $\delta_{i,h}$  denote the phase angle of the voltage in bus $i\in\mathcal{N}$ in time slot $h\in\mathcal{H}$. We use vector $\boldsymbol {\delta}_{h}=\left(\delta_{i,h},\, i\in\mathcal{N}\right)$ to denote the profile of voltage phase angles in all buses in time slot $h$. We define the vector of decision variables $\psi^{\text{ISO}}=(\boldsymbol{p}^{\text{conv}}_j,\boldsymbol{p}^{\text{ren}}_j,  \boldsymbol{\eta}^{\text{ISO}}_{j,h}, \alpha^{\text{ISO}}_{j,h}, \,\boldsymbol{x}_i,\,\boldsymbol {\delta}_{h},h\in\mathcal{H},j\in\mathcal{J},i\in\mathcal{N})$ for the ISO.
The ISO's objective function~is
\begin{align}\label{EPAR}
f^{\text{ISO}}\left(\psi^{\text{ISO}}\right)= \sum_{h\in\mathcal{H}}\Big(&\sum_{j\in\mathcal{M}} \big(\!C^{\text{conv}}_{j,h}(p_{j,h})+\!\vartheta^{\text{c}}\widetilde{\mathcal{C}}_{j,h}^{\beta^{\text{ISO}}_j}\left(\boldsymbol{\eta}^{\text{ISO}}_{j,h}, \alpha^{\text{ISO}}_{j,h}\right)\!\!\big) \!\nonumber\\& +\sum_{i\in\mathcal{N}} \Omega_{i,h}(x_{i,h})\Big),
\end{align}
\noindent where $\vartheta^{\text{c}}$  is a positive weighting coefficient.
Minimizing $f^{\text{ISO}}\left(\psi^{\text{ISO}}\right)$ will enable timely adjustment of control settings to jointly reduce  the consumers' discomfort cost and the generators' generation cost, as well reducing the likelihood of high generation shortage. Thus, it can improve the economic efficiency of the system operation.   We formulate the ISO's centralized  problem as
 \begin{subequations}  \label{ISO_problem}
\begin{align}\label{ISO_obj}
&\displaystyle \underset{\psi^{\text{ISO}}} {\text{minimize}}\,\,\,f^{\text{ISO}}\left(\psi^{\text{ISO}}\right)\\
&\text{subject to constraints (\ref{demand}) and (\ref{gen1})},\;i\in\mathcal{N},\,j\in\mathcal{M},\label{ISO0}\\
&-l_{i,h}=\!\!\sum_{(i,r)\in\mathcal{L}}\!\!\!\!b_{i,r}(\delta_{i,h}-\delta_{r,h}),\hspace{1.1cm} i\in\mathcal{N}, \, h\in\mathcal{H},\label{ISO1}
\\
&p^{\text{conv}}_{j,h}+p^{\text{ren}}_{j,h}=\!\!\sum_{(j,r)\in\mathcal{L}}\!\!\!\!b_{j,r}(\delta_{j,h}-\delta_{r,h}),\hspace{0.25cm}  j\in\mathcal{M}, \, h\in\mathcal{H},\label{ISO2}\\
&\big|b_{r,s}(\delta_{r,h}-\delta_{s,h})\big|\leq p_{r,s}^{\text{max}} ,\hspace{1.15cm}   (r,s)\in \mathcal{L},\;\,h\in\mathcal{H},\label{ISO3}\\
 &\hspace{-0cm} p^{\text{ren}}_{j,h}-p^{\text{ren},k}_{j,h}-\alpha^{\text{ISO}}_{j,h}\leq\eta_{j,h}^{\text{ISO},k},\,j\in\mathcal{M},\,k\in\mathcal{K}, \,h\in\mathcal{H},\label{ISO4}\\
 & 0\leq \alpha_{j,h}^{\text{ISO},k}, \eta_{j,h}^{\text{ISO},k},\hspace{2.8cm}  j\in\mathcal{M}, \, h\in\mathcal{H}.\label{ISO5}
\end{align}
\end{subequations}
Constraints (\ref{ISO1}) and (\ref{ISO2}) are the nodal power balance equations. Constraint (\ref{ISO3}) is the  line flow limit, where $p_{r,s}^{\text{max}}$ is the maximum power flow limit for line $(r,s)\in\mathcal{L}$ and $b_{r,s}$ is the admittance of line $(r,s)\in\mathcal{L}$.  Constraint (\ref{ISO4}) is for auxiliary variable $\eta^{\text{ISO},k}_{j,h}$ corresponding to  term $ [p^{\text{ren}}_{j,h}- \widehat{p}^{\,\text{ren}}_{j,h}]^{+}$.

Problem (\ref{ISO_problem}) is an optimization problem with a convex objective function (remind that CVaR is convex) and linear  constraints. By considering one bus as a slack bus (e.g., bus $1$ with $\delta_{1,h}=0, \,h\in\mathcal{H}$), we have the following theorem.

\vspace{1mm}
\noindent\textbf{Theorem 1} \textit{If  the  ISO's centralized problem in (\ref{ISO_problem}) has a feasible point, then the optimal solution  exists and is unique.}
\vspace{1mm}

To solve  the centralized problem in (\ref{ISO_problem}), the ISO needs  the information about the load aggregators' discomfort cost, conventional generators'  cost, and renewable units'  historical data samples. However, these information may not be available to the ISO. Also, solving problem (\ref{ISO_problem}) can be  computationally difficult in large networks.
Instead, we develop a decentralized algorithm. We show that the ISO can determine the  price signals $\rho_{b,h}$ for bus $b\in\mathcal{N}\cup\mathcal{M}$ and penalties $\theta_{j,h}$, $h \in \mathcal{H}$ for generators $j\in\mathcal{M}$, such that the competitive equilibrium of the market coincides with the \text{unique solution} of the ISO's centralized problem in (\ref{ISO_problem}). Note that the energy market equilibrium corresponds to the optimal solution to problems (\ref{agg_prob}) and (\ref{util_problem}) for all entities.

\vspace{-0mm}
\section{Energy Market Interactions}\label{s3}

The ISO has no\textit{ direct}
control over the generators' and load aggregators' behavior. Alternatively, the ISO provides the entities with sufficient access to the day-ahead energy market to determine their own generation and load demand. The entities compete with each other to optimize  their local problems (\ref{agg_prob}) and (\ref{util_problem}). The decision vector of the load aggregator $i$ is the controllable load profile  $\boldsymbol{x}_i$ and the decision vector for generator $j$ is ${\psi}_j$. The ISO   influences 
the entities by using the nodal prices $\boldsymbol{\rho}=(\boldsymbol{\rho}_{b},\;b\in\mathcal{N}\cup \mathcal{M})$  and the penalties $\boldsymbol{\theta}=(\boldsymbol{\theta}_j,\,j\in\mathcal{M})$ for   renewable generators.   The goal of the ISO is to determine  vectors $\boldsymbol{\rho}$ and $\boldsymbol{\theta}$, such that the  market equilibrium coincides with the optimal solution of the centralized problem (\ref{ISO_problem}).  
The idea is to formulate the \textit{Lagrange relaxation} of the ISO's problem (\ref{ISO_problem}) in order to divide problem (\ref{ISO_problem})  into several subproblems. Then, we determine vectors $\boldsymbol{\rho}$ and $\boldsymbol{\theta}$ such that the subproblems becomes the same as load aggregators and generators problems  in (\ref{agg_prob}) and (\ref{util_problem}). This approach is viable since Due to
 problem (\ref{ISO_problem}) is convex and the constraints are linear. Thus,  the strong duality gap condition (Slater’s condition) is satisfied
if a feasible solution exists \cite[Ch. 5]{boyd2004convex}.

  Let $\lambda^{\text{agg}}_{i,h},\,i\in\mathcal{N},\, h\in\mathcal{H}$ denote the Lagrange multiplier associated with the equality constraint (\ref{ISO1}). Let $\lambda^{\text{gen}}_{j,h},\, h\in\mathcal{H}, j\in\mathcal{M}$ denote the Lagrange multiplier associated with the equality constraint (\ref{ISO2}). Let $\overline{\mu}_{r,s,h}$ and $\underline{\mu}_{r,s,h},\, h\in\mathcal{H}, \;(r,s)\in\mathcal{L}$ denote the Lagrange multiplier associated with the upper and lower inequalities (\ref{ISO3}), respectively. Also, let $\gamma^{\text{ISO},k}_{j,h},\, j\in\mathcal{M},\,k\in\mathcal{K}, \,h\in\mathcal{H}$ denote the Lagrange multiplier associated with constraint (\ref{ISO4}). 
We define  vector $\phi^{\text{ISO}}=(\lambda^{\text{agg}}_{i,h},\,\lambda^{\text{gen}}_{j,h},\,\overline{\mu}_{r,s,h},\,\underline{\mu}_{r,s,h},\gamma^{\text{ISO},k}_{j,h},\,i\in\mathcal{N}, $ $ j\in\mathcal{M}, (r,s)\in \mathcal{L},\, k\in\mathcal
K,\, h\in\mathcal{H})$.  We have the following main result.
 
 \vspace{1mm}
\noindent\textbf{Theorem 2} \textit{The equilibrium of the energy market coincides with the unique solution to the  ISO's centralized problem in (\ref{ISO_problem}) if and only  if for $i\in\mathcal{N},j\in\mathcal{M},\, h\in\mathcal{H}$ the ISO sets }
\begin{subequations}
\begin{align}
&\rho_{i,h}=-\lambda^{\text{agg}}_{i,h},\\
&\rho_{j,h}=\lambda^{\text{gen}}_{j,h},\label{penal0}\\
& \theta_{j,h}=\frac{\sum_{k\in\mathcal{K}}\gamma_{j,h}^{\text{ISO},k}}{1-\vartheta^c+\sum_{k\in\mathcal{K}}\gamma_{j,h}^{\text{ISO},k}}\Big(\frac{1-\beta_j}{1-\beta^{\text{ISO}}_j}\Big).\label{penal}
\end{align}
\end{subequations}

%
\noindent The proof can be found in Appendix A.  It suggests a decentralized algorithm to determine the solution to problem (\ref{ISO_problem}).

 In a \text{decentralized} algorithm,   the load aggregators, generators and ISO interact with each other.   Let $q$ denote the iteration index. Our algorithm involves the
\textit{initiation phase} and \textit{ market trading phase}.  Algorithm 1 describes the interactions among the load aggregators, generators, and ISO. Algorithm 1 is based on the projected gradient method. The step size in iteration $q$ of this algorithm is denoted by $\epsilon^{q}$.

\textit{Initiation phase:} Lines 1 to 4 describe the initiation phase. 

\textit{Market trading phase:}  The loop involving  Lines 5 to 12 describes the market trading phase, in which  the load aggregators, generators and ISO update their decision variables in an iterative fashion.  This phase includes the following parts:

 \textit{ 1) Information exchange:}  In Line 6, each load aggregator  $i$ sends its  load profile $(l^q_{i,h}=l^{\text{c},q}_{i,h}+l^{\text{b}}_{i,h},\,h\in\mathcal{H})$  to the ISO via the communication network. Each generator $j$ sends the generation profile $\boldsymbol{p}^{\text{conv},q}_{j}$ and $\boldsymbol{p}^{\text{ren},q}_{j}$ to the ISO as well.

\textit{2) ISO's update:} In Line 7, when the ISO receives the information from the entities, it obtains the updated values of the voltage angles $\delta_{b,h}^{q+1}$ for  $b\in\mathcal{N}\cup\mathcal{M}$ in time slot $h\in\mathcal{H}$ and the vector of Lagrange multipliers   $\boldsymbol{\phi}^{\text{ISO},q+1}$ using the gradient of the Lagrangian function $f_{\text{Lag}}^{\text{ISO}}(\psi^{\text{ISO}},\phi^{\text{ISO}})$ (See (S-1) in Appendix A) in iteration $q$ as follows:
\begin{subequations}
\begin{align}
     	&\delta_{b,h}^{q+1}=\delta_{b,h}^{q}+\epsilon^q\nabla_{\delta_{i,h}^{q}}\,f_{\text{Lag}}^{\text{ISO}}(\psi^{\text{ISO},q},\phi^{\text{ISO},q}),\label{signal1}\\
     	&{\phi}^{\text{ISO},q+1}={\phi}^{\text{ISO},q}\!+\!\epsilon^q\nabla_{\!\!\boldsymbol{\phi}^{\text{ISO},q}}\!f_{\text{Lag}}^{\text{ISO}}(\psi^{\text{ISO},q},\phi^{\text{ISO},q}),  \label{signal2}
     	   \end{align}
\end{subequations}
where $\nabla$ is the gradient operator. The ISO uses the results of Theorem 2 to compute the updated values of the  nodal prices $\boldsymbol{\rho}^{q+1}$ and penalties $\boldsymbol{\theta}^{q+1}$.
In Line 8, the ISO communicates the above control signals  to the corresponding load aggregator and  generator.

\begin{algorithm}[t]\label{al1}\small
 	\caption{ Decentralized Energy Market Trading Algorithm.}
 	\begin{algorithmic} [1]
 	
 	\STATE  Set $q:= 1$ and $\xi:=10^{-2}$.
 	\STATE Each load aggregator  $i\in\mathcal{N}$ randomly initializes its controllable load profile $\boldsymbol{x}^{1}_i$. 
 	\STATE Each generator $j\in\mathcal{M}$ randomly initializes its generation profile $\boldsymbol{p}_j^{1}$.  Each generator $j$ with renewable units  initializes vectors $\boldsymbol{\alpha}_j^1$ and  $\boldsymbol{\eta}_j^1$ randomly, and provides the ISO with  confidence level $\beta_j$.
 	\STATE The ISO  initializes  the voltage angles $\delta_b^1,\,b\in\mathcal{N}\cup\mathcal{M}$ and the vector of Lagrange multipliers   $\boldsymbol{\phi}^{\text{ISO},1}$.

\STATE \textbf{Repeat}
\STATE \hspace{1mm} Each load aggregator $i$ and generator $j$, sends $l^q_{i,h},\,h\in\mathcal{H}$ and   \\ \hspace{1mm} $\boldsymbol{p}^{\text{conv},q}_{j}$  and $\boldsymbol{p}^{\text{ren},q}_{j}$  to the ISO, respectively.
\STATE \hspace{1mm} ISO updates the voltage angles $\delta_{b,h}^{q},\,h\in\mathcal{H},b\in\mathcal{N}\cup\mathcal{M},$ and \\ \hspace{1mm} the   vector of Lagrange multipliers   $\boldsymbol{\phi}^{\text{ISO},q}$ according to (\ref{signal1}) \\ \hspace{1mm} and (\ref{signal2}).
 	
 	\STATE \hspace{1mm} ISO communicates the updated nodal price $\boldsymbol{\rho}^{q+1}$ and penalties  \\ \hspace{1mm}  $\boldsymbol{\theta}^{q+1}$ to the entities.
 	\STATE \hspace{1mm} Each load aggregator $i$ updates the  controllable load profile  $\bm{x}_i^{q}$  \\ \hspace{1mm}  by solving its local problem in (3).
 	\STATE \hspace{1mm} Each generator $j$  updates its  decision vector $\psi_j^{q}$  by solving its \\ \hspace{1mm} local  problem in (10).
 		\STATE \hspace{1mm} $q:=q+1$. The step size $\epsilon^q$ is updated.
 	
 	\STATE \textbf{Until}   $||\delta_{b,h}^{q}-\delta_{b,h}^{q-1}||\leq \xi,\,\,h\in\mathcal{H},b\in\mathcal{N}\cup\mathcal{M}$.
 	 	\end{algorithmic} 
 \end{algorithm} 
 \normalsize

 \textit{ 3) Load aggregator's update:}  When load aggregator  $i$ receives the control signals $\rho_{i,h}^{q+1},\,h\in\mathcal{H}$ from the ISO, in Line 9, it updates its controllable load profile $\bm{x}_i^{q}$ by solving its local problem in (3). It is a quadratic program with linear constraints and can be solved efficiently by the load aggregator.

 \textit{Generator's update:}  When generator $j$ receives the control signals $\rho_{j,h}^{q+1}$ and $\theta^{q+1}_{j,h},\,h\in\mathcal{H}$  from the ISO, in Line 10, it updates its  decision vector $\psi_j^{q}$  by solving its local problem in (10). It is a quadratic program with linear constraints and can be solved efficiently by the generator using its local information about its conventional and renewable units.

 \textit{ Step size update:} We use a nonsummable diminishing step size with conditions $\lim_{q\rightarrow \infty} \epsilon^q=0$, $ \sum_{q=1}^{\infty}{\epsilon^q}= \infty$, and $ \sum_{q=1}^{\infty}{\big(\epsilon^q\big)^2}= \infty$. One example is $\epsilon^q=\frac{1}{a+b q}$, where $a$ and $b$ are positive constant coefficients.  In Line 11, the step size is updated. In Line 12, the stopping criterion is given. For stoppin criterion, we have used the convergence of the voltage angles, since they depends on the generation and load level at all buses. Hence, the convergence of the voltage angles implies the convergence of all generators' and load aggregators' decision variables. The proposed Lagrange relaxation-based algorithm converges for a nonsummable diminishing step size~\cite[Ch. 5]{boyd2004convex}.




\vspace{-0in}
\section{Performance Evaluation}\label{s5}
In this section, we evaluate the performance of our proposed decentralized algorithm on  an IEEE 30-bus test system with $6$ generators and 21 load aggregators.

\textit{1) Simulation setup:} The data for the test system and the generators cost functions are given in \cite{ieee1}. Since buses $2$, $5$, and $8$ have both generator and load aggregator in the original test system, we add the new virtual buses $31$, $32$, and $33$ for the load aggregators as shown in Fig. S-1 in Appendix B.   The trading horizon is one day with $H = 24$
one-hour time slots.  
The weight coefficient  in the ISO's objective function (\ref{EPAR}) is set to $\vartheta^{\text{c}}=2$ (without units). 

 To obtain different baseload patterns for the load aggregators, we use a load pattern for about 5 million consumers (which includes residential, commercial, and industrial consumers) from Ontario, Canada power grid database \cite{ont} from November 1 to November 21 2016. We  scale the  load pattern for each bus, such that the average baseload becomes equal to  $60\%$ of the load demand of that bus  in \cite{ieee1}. To simulate the controllable load demand of each bus, we randomly generate the desirable load profile of $500$ to $1000$ controllable loads of types 1 and 2 with average demand of $2$ to $15$  kW at each bus.   Limits $x_{a,i,h}^{\text{max}}$ and $x_{a,i,h}^{\text{min}}$ for controllable load $a$ in bus $i$ are set to $\pm 30\%$ of the desirable demand of that load in time slot $h$. Limits $X_{a,i}^{\text{max}}$ and $X_{a,i}^{\text{min}}$ for controllable load $a$ in bus $i$ are set to $\pm 5\%$ of the desirable total demand of that load.  The discomfort coefficients $\omega_{a,i}$ for controllable load $a$ of type 1 are  randomly chosen from  a truncated
normal distribution, which is lower bounded by zero and has a mean value of $\omega_i^{\text{avg}}=15\, \text{cents}/(\text{kWh})^2$ and a standard deviation of $5 \, \text{cents}/(\text{kWh})^2$. The discomfort coefficients $\omega_{a,i,h},\, h\in\mathcal{H}$ for controllable load $a$ of type 2 in bus $i$   are  randomly chosen from  a truncated
normal distribution, which is lower bounded by zero and has a mean value of $\omega_i^{\text{avg}}=15\, \text{cents}/(\text{kWh})^2$ and a standard deviation of $5 \,\text{cents}/(\text{kWh})^2$. The discomfort coefficients $\omega'_{a,i,h},\, h\in\mathcal{H}$ for controllable load $a$ of type 2 in bus $i$ are set to $50\, \text{cents}/(\text{kWh})^2$. 

We assume that generators in buses $8$ and $11$ have PV panel and wind turbine, respectively, in addition to their conventional units. To obtain the samples for   the output
power of the wind turbine and PV panel, we use the available historical
 data from Ontario, Canada power grid database \cite{ont}, 
from November 1 to November 21 2016. For each renewable generator, we scale down the
available historical data, such that the average output power
 over the historical data becomes
equal to 4 MW. 

The step size is $\epsilon^{q}=\frac{1}{10+0.2q}$. For the benchmark scenario, we consider
a system without renewable units and DR program.  We perform
simulations using Matlab R2016b in a PC
with processor Intel(R) Core(TM) i7-3770K CPU@3.5~GHz.  



\begin{figure}[t]
\centering
\includegraphics[width=3.55in]{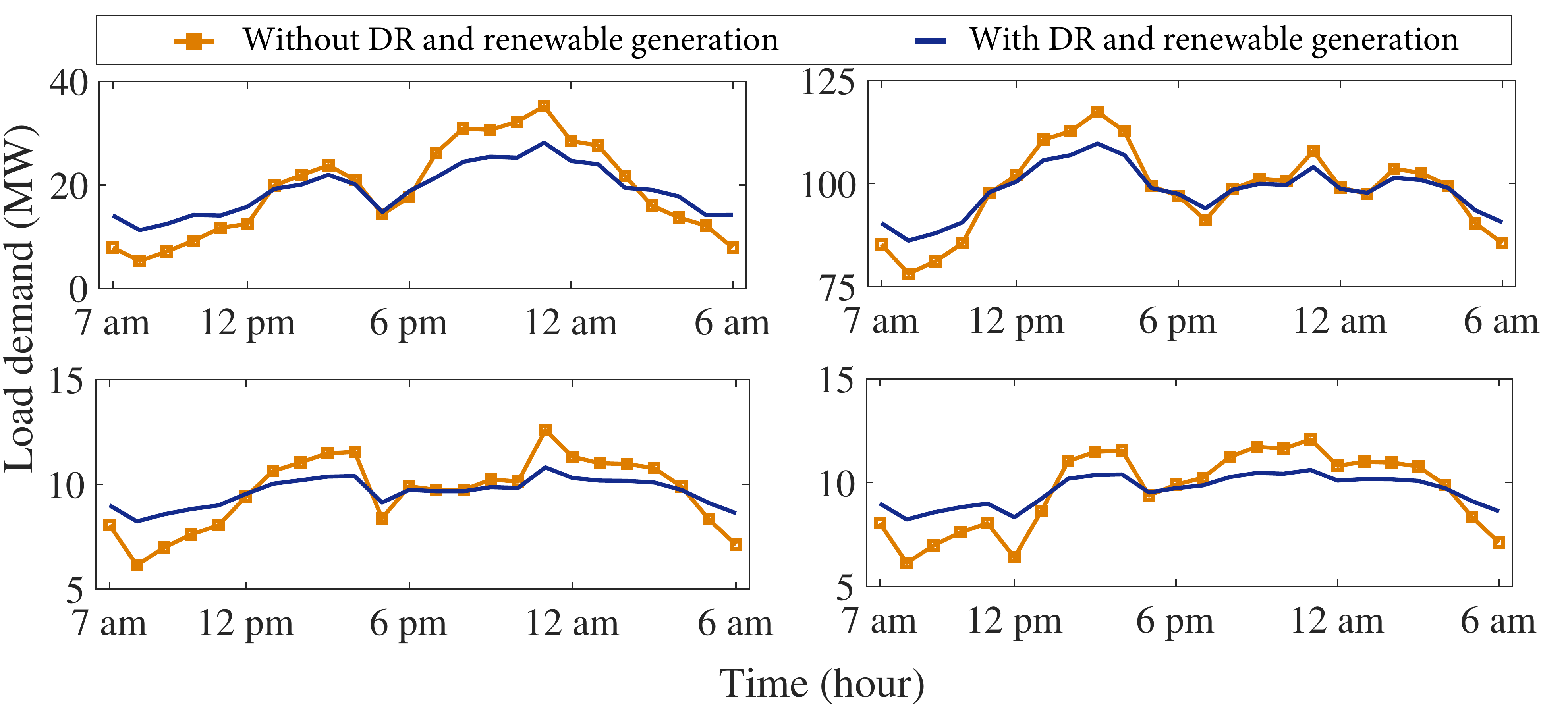}
\vspace{-7mm}
\caption{Load demand profiles over $24$ hours in buses $31$ (up-left), $32$ (up-right), $17$ (down-left), and $30$ (down-right) with and without DR and renewable generators.}
\vspace{2mm}
\label{totload}
\centering
\includegraphics[width=3.55in]{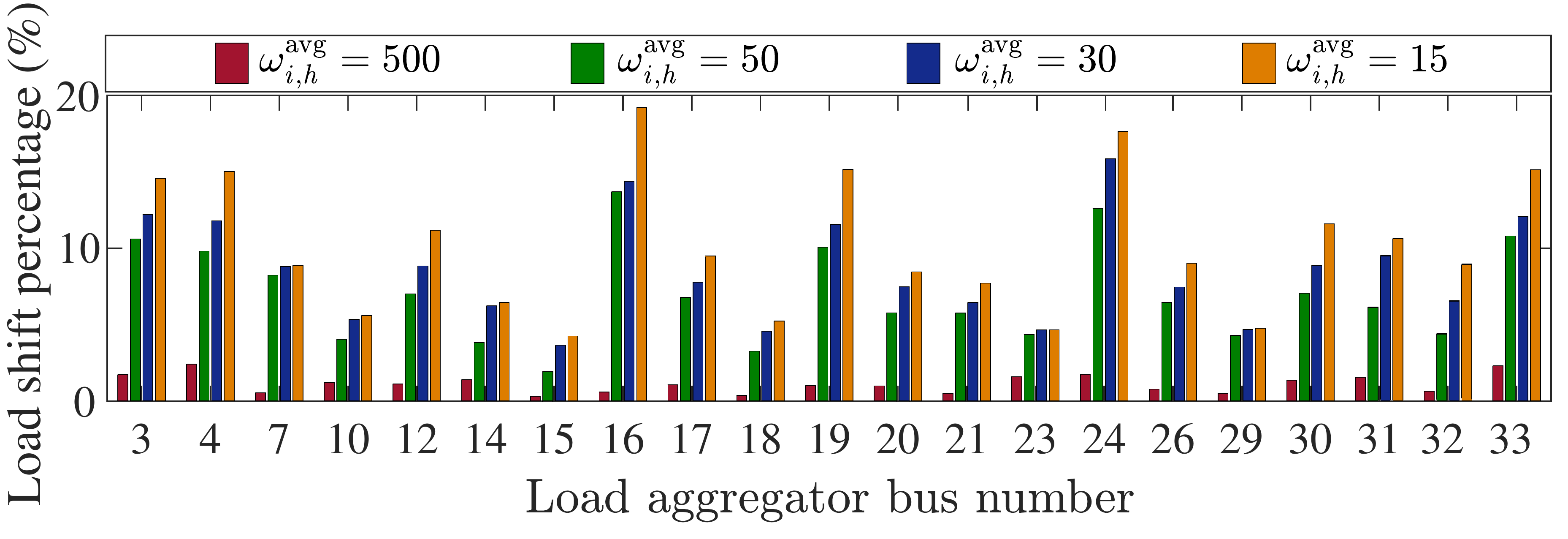}
\vspace{-7mm}
\caption{The values of shift load percentage for different load aggregators with $\omega_{i}^{\text{avg}}=500,\,50,\,30,\,$ and $15$ cents/(kWh)$^2$.}
\label{shiftload}
\vspace{2mm}
\centering
\includegraphics[width=3.55in]{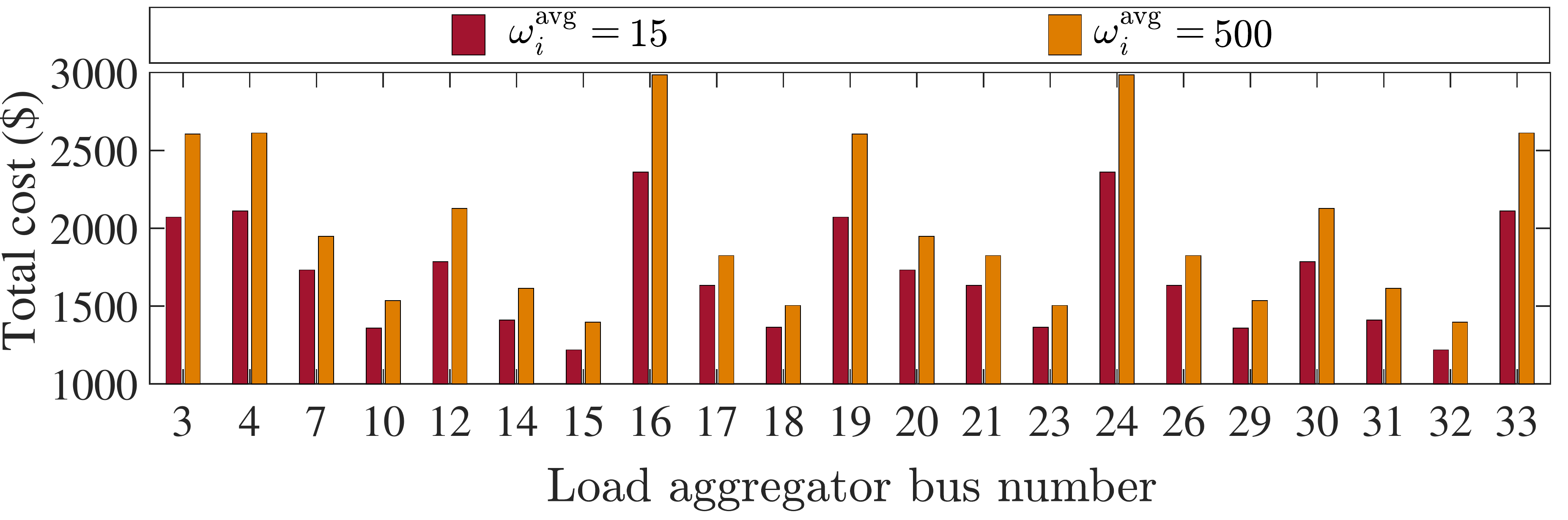}
\vspace{-7mm}
\caption{The load aggregator's cost with low and high discomfort cost's scaling coefficients.}
\label{aggcost}
\vspace{2mm}
\centering
\includegraphics[width=3.55in]{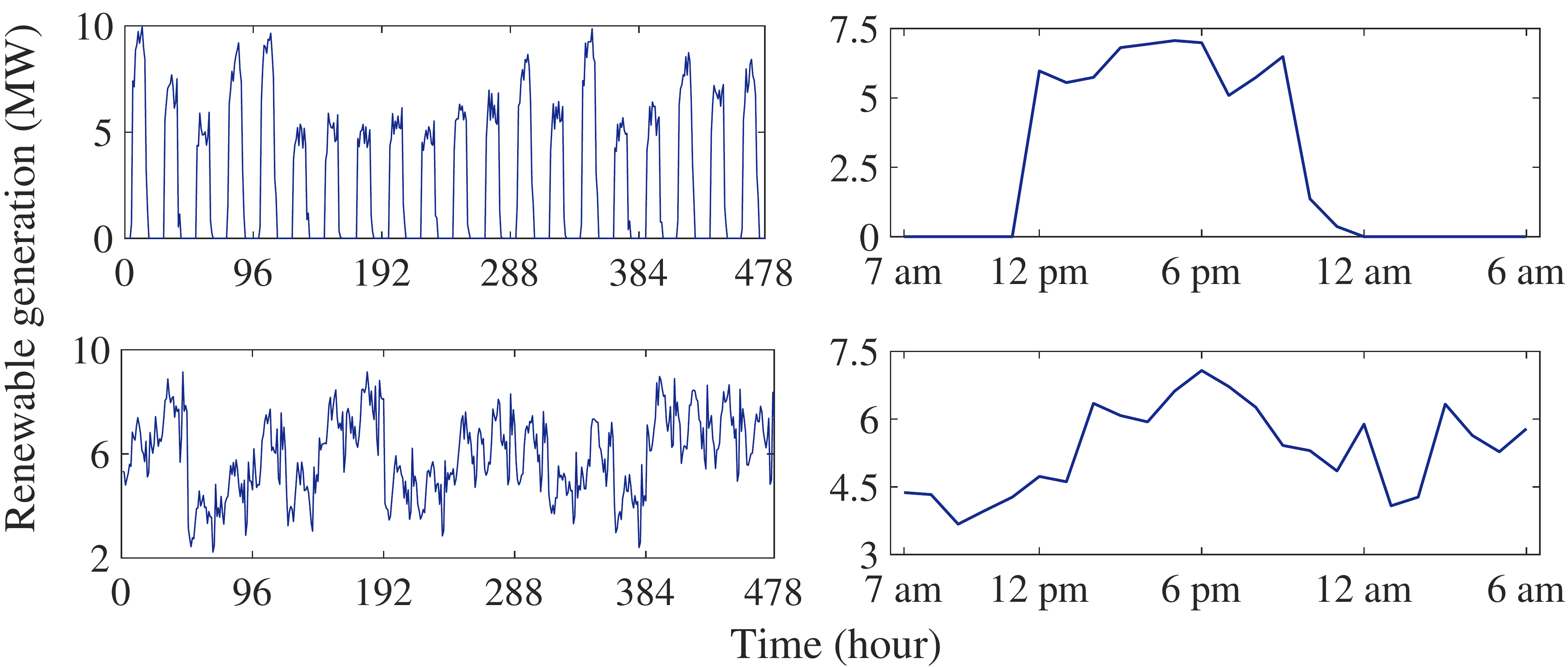}
\vspace{-7mm}
\caption{(left figures) The PV panel and wind turbine historical data samples. (right figures) The presumed output power of the PV panel and wind turbine in buses $11$ and $13$. }
\label{renew}
\vspace{-4mm}
\end{figure}

\textit{2) Load aggregators:} Each  load aggregator executes Algorithm 1  to modify the controllable load demand of its users. Fig. \ref{totload} shows the load profile of the load aggregators in buses  $17$, $30$, $31$ (the load in bus $2$), $33$ (the load in bus $5$)    and the benchmark system without renewable generators and DR program. Peak shaving can be observed in the load profiles. Results for all load aggregators verify that by executing the decentralized Algorithm 1, the peak load demand is reduced by $16.5\%$ on average.  Since the peak load may occur in different times for different load aggregators, we consider the index of \textit{load shift percentage}, which is the ratio of the total shifted load demand to the aggregate  demand over $24$ hours.  In Fig. \ref{shiftload}, the shift load percentage index for different values of the average scaling coefficient $\omega_{i}^{\text{avg}}$ at all load buses are provided. When  $\omega_{i}^{\text{avg}}$ increases in a bus, the load aggregator modifies a lower amount of controllable loads  due to a higher discomfort cost of its users, and thereby the load shift percentage index  decreases.  Load shifting is performed with the goal of reducing the total cost in (\ref{costt}). Fig. \ref{aggcost} shows that the  total cost in (\ref{costt}) of the load aggregators is lower (by about $18\%$) when the scaling coefficients $\omega_{i}^{\text{avg}}$ are lower, since the load aggregators can benefit from the price fluctuations and shift more amount of load demand to the hours with lower price values. 
\begin{figure}[t]
\centering
\includegraphics[width=3.55in]{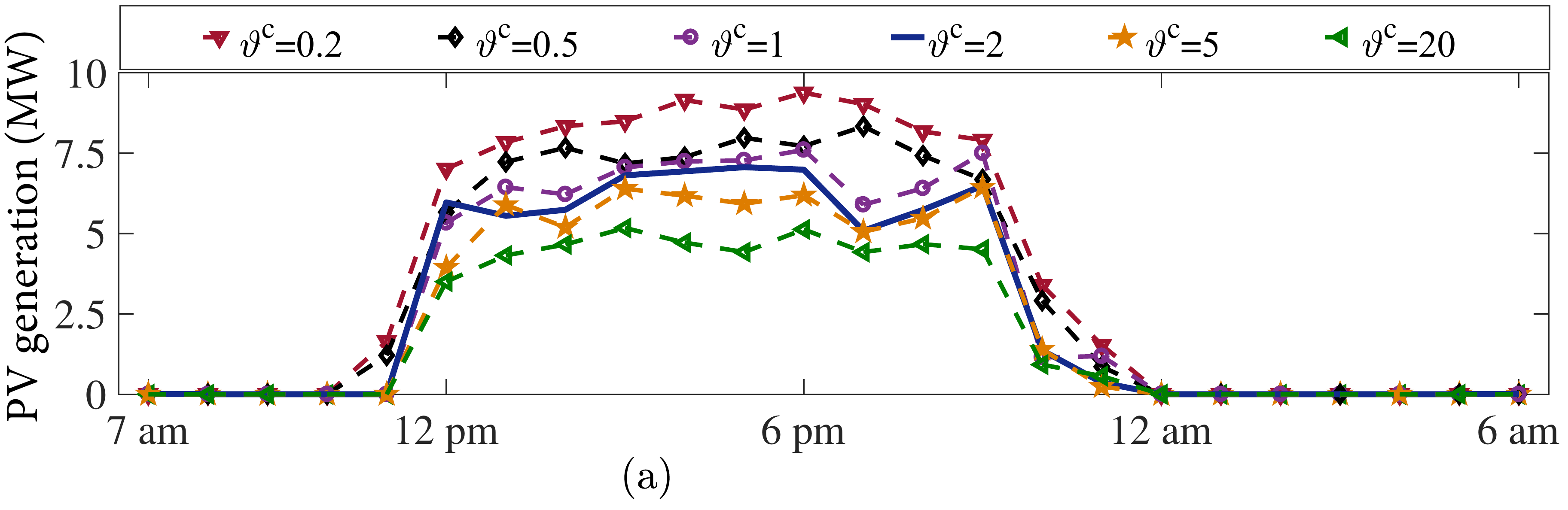}
\includegraphics[width=3.55in]{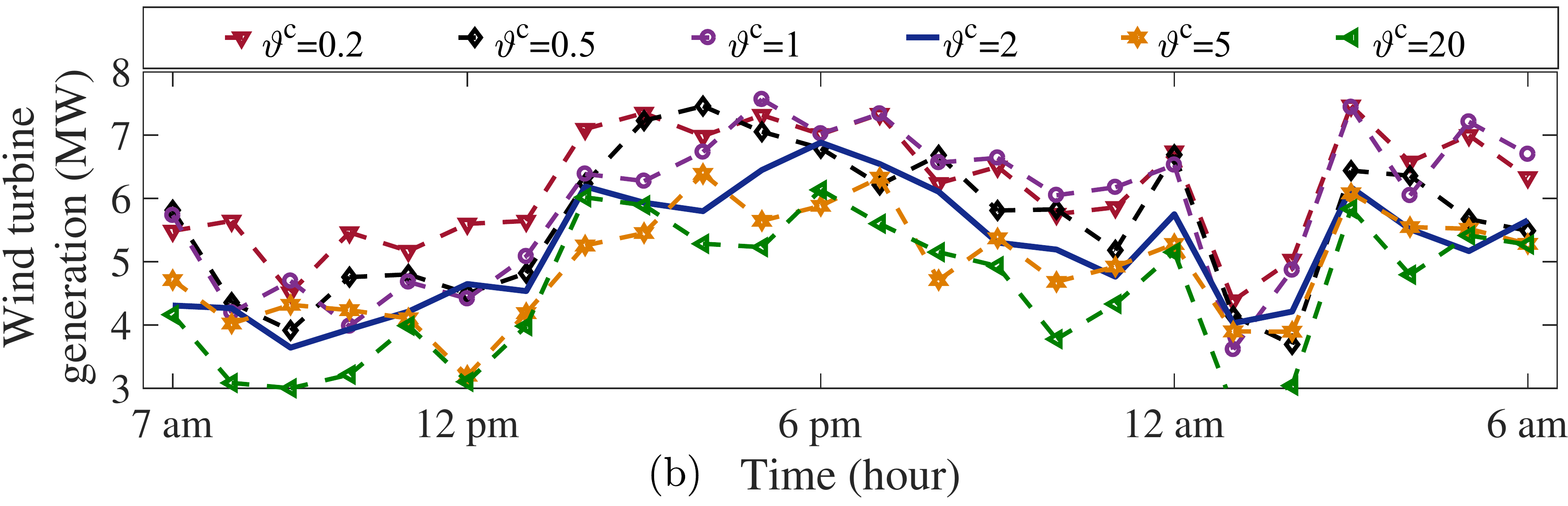}
\vspace{-7mm}
\caption{The presumed output power of ($a$) the PV panel in bus $11$, and ($b$) the wind turbine in bus $13$ for different values of coefficient $\vartheta^{\text{c}}$.}
\label{renew1}
\vspace{4mm}
\centering
\includegraphics[width=3.55in]{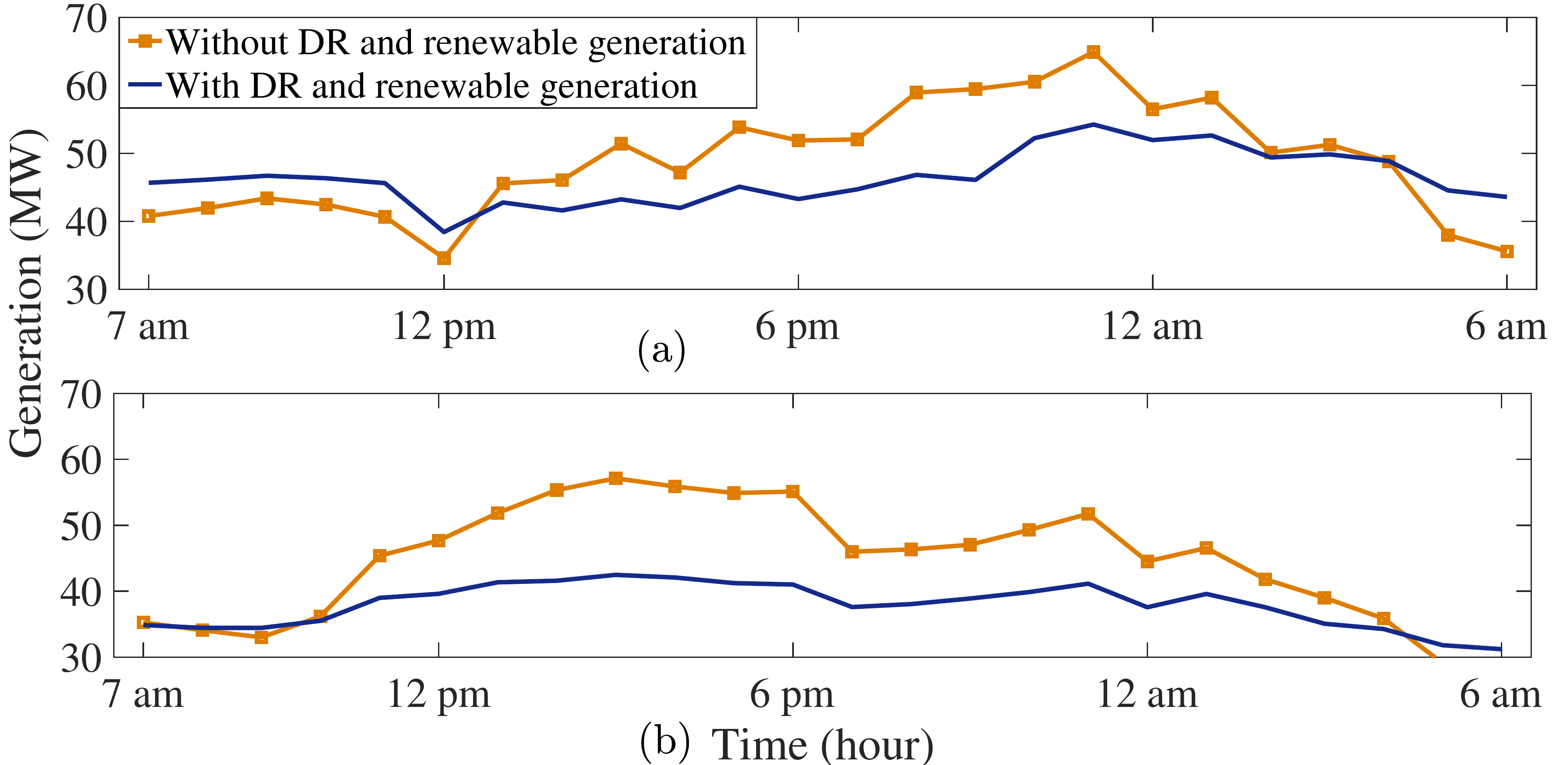}
\vspace{-7mm}
\caption{The  generation profile of the generators ($a$) $11$  and ($b$) $13$. }
\label{gen}
\vspace{2mm}
\centering
\includegraphics[width=3.54in]{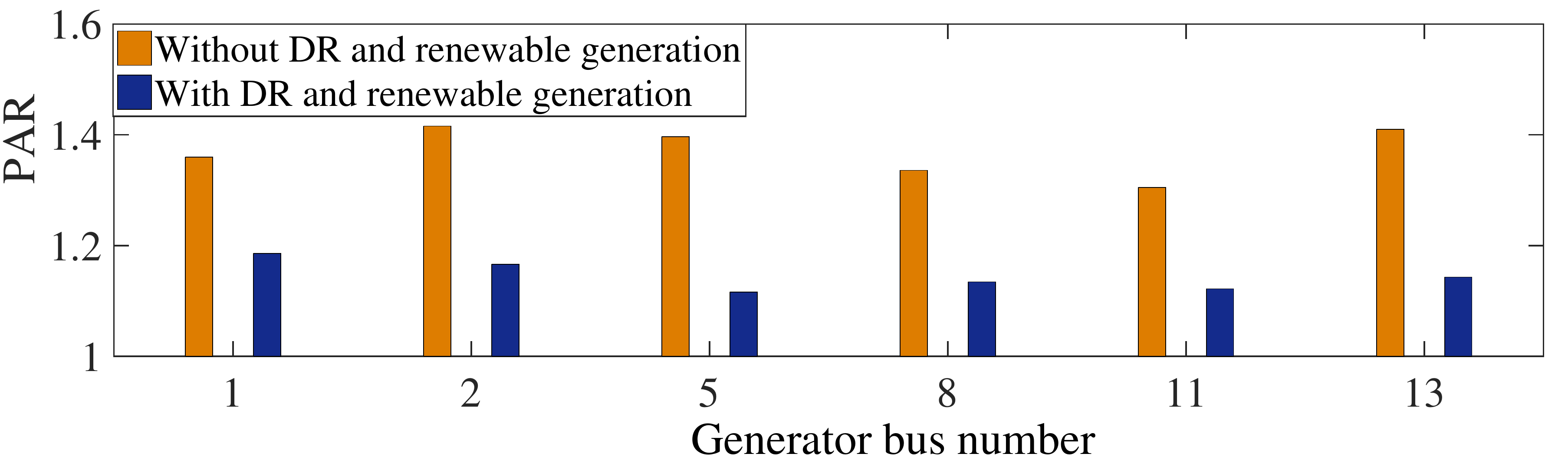}
\vspace{-7mm}
\caption{The PAR in the generation of the generators with and without DR and renewable generation.}
\label{parg}
\vspace{-3mm}
\end{figure}

\textit{3) Renewable generators:}  Generator $11$ has a PV panel and generator $13$ has a wind turbine in addition to their conventional units. Recall that these generators sell the  \textit{presumed} output power of the renewable units in the day-ahead market, and pay a penalty for generation shortage. In Algorithm 1, the generators use the historical data of the renewable resources and respond to the control signals (\ref{penal0}) and (\ref{penal}) to determine the amount of power that they plan to sell in the market.  Fig. \ref{renew} shows the historical data and the presumed output power of the PV panel in bus $11$ and the wind turbine in bus $13$ for the confidence levels $\beta_{11}=\beta_{13}=0.9$. The obtained presumed output power will have the lowest risk in the day-ahead market for these generators.
Two parameters affect the presumed output power of the renewable generator: the value of weight coefficient $\vartheta^{\text{c}}$ in the ISO's objective function (\ref{EPAR}) and the values of confidence levels $\beta_j$ (set by  generator $j$) and $\beta_j^{\text{ISO}}$ (set by the ISO). The penalty $\theta_{j,h}$ in (\ref{penal}) for the generation shortage increases when $\vartheta^{\text{c}}$ increases. That is, when the risk-averse ISO puts a higher weight on the risk of generation shortage, it assigns a larger penalty. Fig. \ref{renew1} $(a)$ (for bus $11$) and  Fig. \ref{renew1} $(b)$ (for bus $13$) show that a larger coefficient $\vartheta^{\text{c}}$  enforces the generators to offer a lower renewable generation  in the market during most of the times. In a similar manner, equation (\ref{penal}) implies that when $\beta_j>\beta_{j}^{\text{ISO}}$, i.e.,  generator $j$ is more risk-averse than the ISO, then the ISO reduces  penalty $\theta_{j,h}$ by factor $\frac{1-\beta_j}{1-\beta^{\text{ISO}}_j}<1$. If $\beta_j<\beta_{j}^{\text{ISO}}$, i.e.,  generator $j$ is more risk-taker than the ISO, then the ISO increases penalty $\theta_{j,h}$ by  $\frac{1-\beta_j}{1-\beta^{\text{ISO}}_j}>1$ to limit the likelihood of  generation shortage.

\begin{figure}[t]
\centering
\includegraphics[width=3.55in]{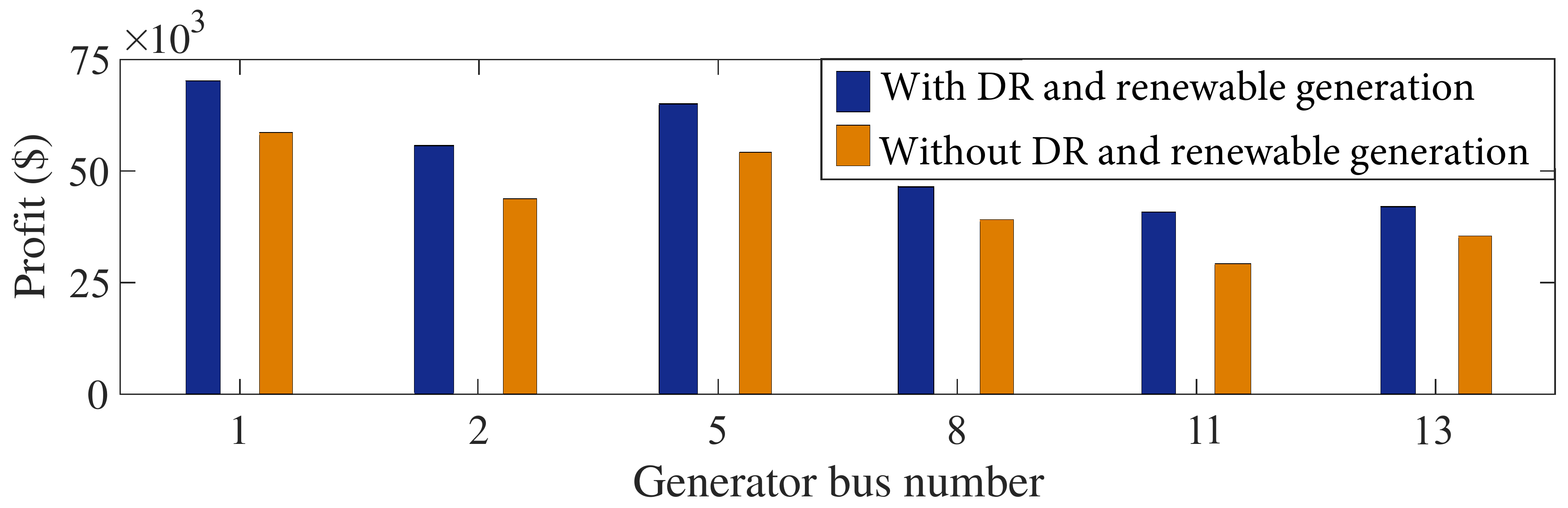}
\vspace{-7mm}
\caption{The profit of the generators with and without DR and renewable generation.}
\label{profit1}
\centering
\vspace{2mm}
\includegraphics[width=3.55in]{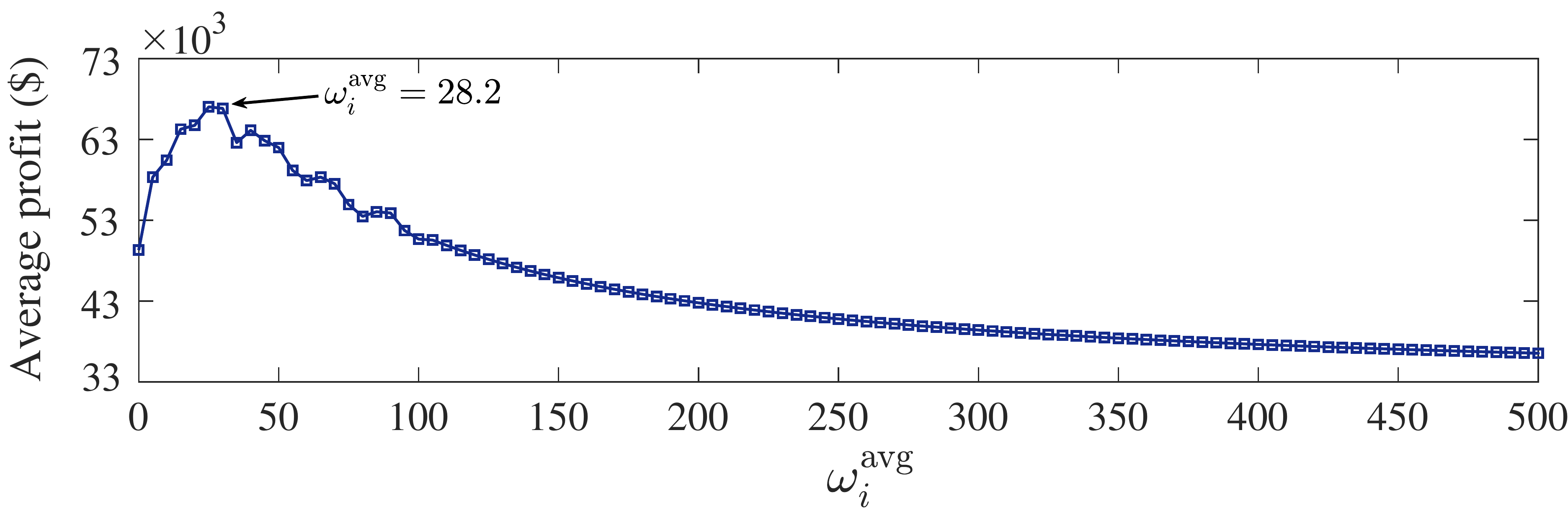}
\vspace{-7mm}
\caption{The average profit of the generators in terms of the average weight coefficients $\omega_{i}^{\text{avg}},\,i\in\mathcal{N}$.}
\label{profit2}
\vspace{-4mm}
\end{figure}

\textit{4) Conventional generators:} The  generators with conventional units can also benefit from the DR program  by reducing the PAR in the aggregate demand. For example, Fig. \ref{gen} shows that  using the DR program makes the generation profile of the conventional units  in bus $11$ and $13$ smoother. In other words, as the demand profiles in different buses  change toward a profile with a lower peak demand, the required power generation follows the same trend.  In order to quantify the impact of the  DR program on the generation profile, we provide the value of the PAR  with and without DR program in Fig. \ref{parg}. Results  show the reduction in PAR   for the generators by $15\%$ on average. 
The reduction in the PAR  can reduce the peak load demand, and thus the generation cost of the generators.  However, the \textit{revenue} of the generators will decrease due to a lower price in the market. Fig. \ref{profit1} shows the profit (revenue minus cost)  for the generators with and without DR program.  The profit of the generators increase by $17.1\%$ on average with DR. To complete the discussion, we  consider different discomfort cost coefficients $\omega_{i}^{\text{avg}},\,i\in\mathcal{N}$ for the load aggregators and report the average profit of generators in Fig. \ref{profit2}. It is interesting that the scenario with the most flexible load demands (i.e., $\omega_{i}^{\text{avg}}=0$) does not lead to the maximum profit for the generators, though the generation cost of the generators are lower in this scenario. The reason is that a higher load flexibility will flatten the load profiles, and thus reduces the price values and  the revenue of the generators. Here, the reduction in the revenue is larger than the reduction  in the costs. The maximum profit is achieved for $\omega_{i}^{\text{avg}}=28.2$ cents/(kWh)$^2$.

\vspace{4mm}
\begin{figure}[t]
\centering
\includegraphics[width=3.55in]{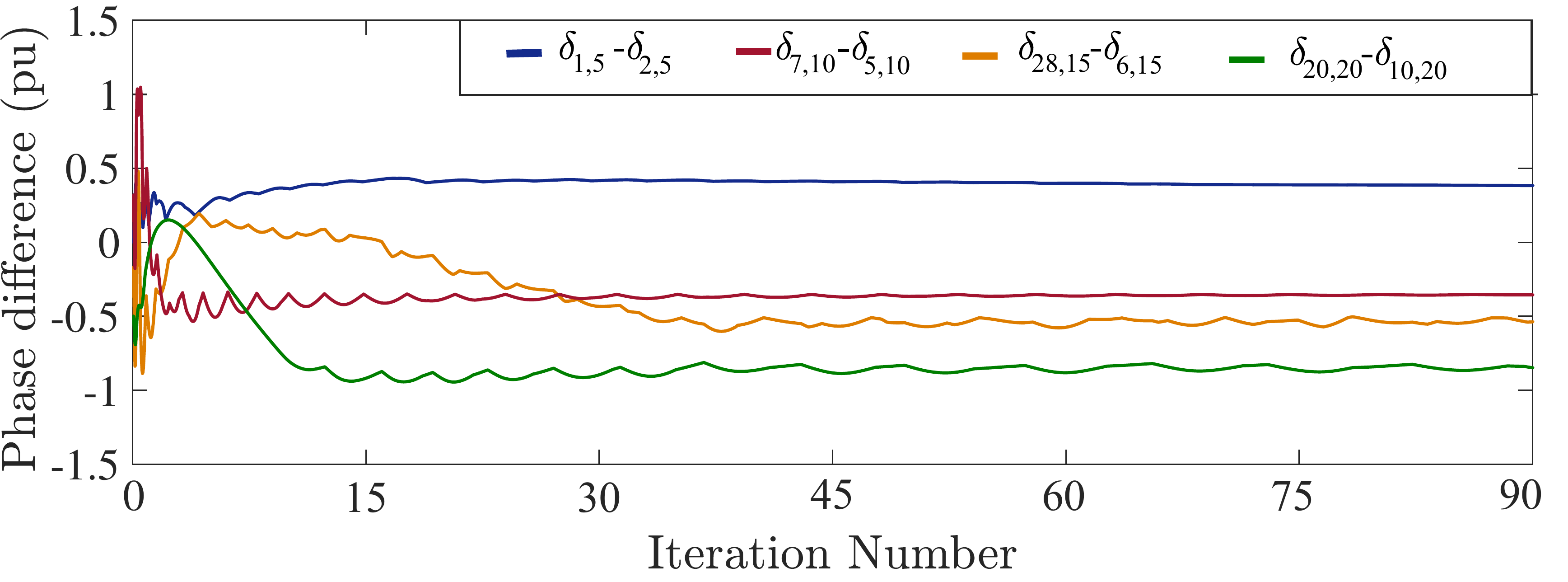}
\vspace{-7mm}
\caption{The convergence of phase difference over transmission lines.}
\label{theta}
\vspace{2mm}
\centering
\includegraphics[width=3.54in]{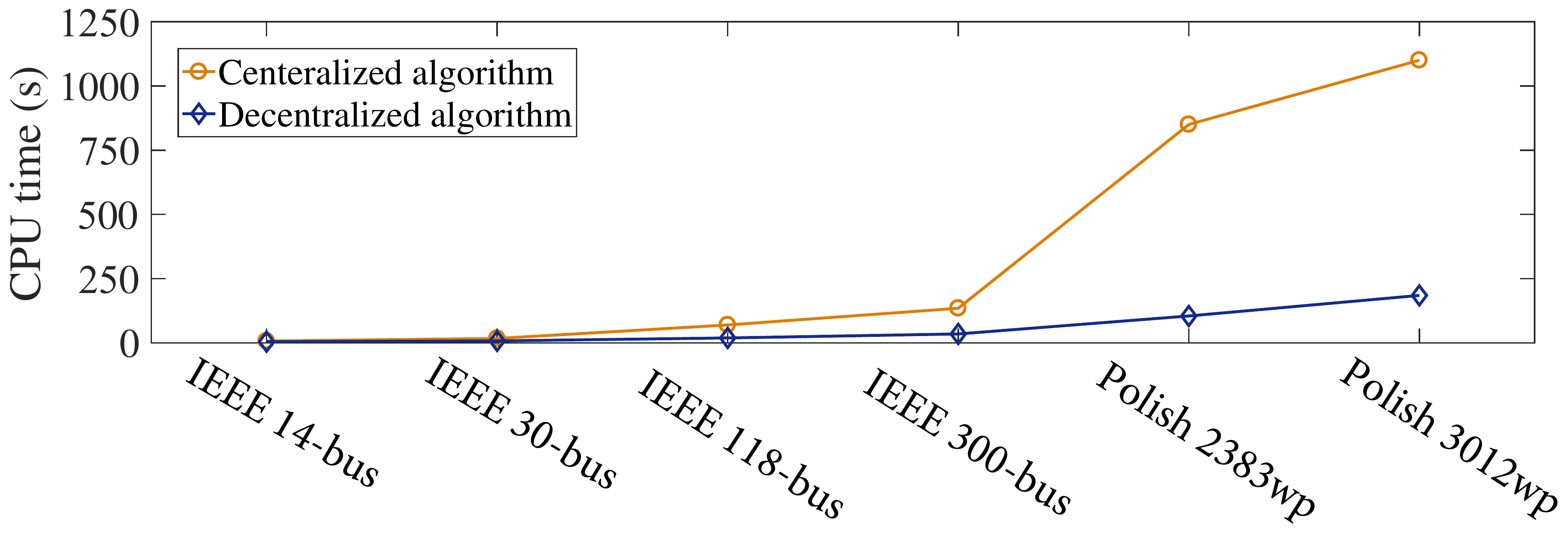}
\vspace{-7mm}
\caption{The CPU time of the centralized and decentralized algorithms.}
\label{run}
\vspace{-4mm}
\end{figure}

\textit{5) Algorithm convergence:}  We study the required number of iterations for convergence, which can be interpreted as an indicator of the number of  message exchanges among the load aggregators, generators and the ISO.   The value of voltage angles depends on all generators' and load aggregators' decision variables. Thus, the convergence of the voltage angles is a viable indicator of the convergence of the decision variables in all buses of the system.  Since the values of the voltage angles of all buses can be added by a constant, we illustrate the convergence of the phase angle difference between the voltages of the buses at the end nodes of  the lines. As an example, we provide the convergence of $\delta_{1,5}-\delta_{2,5}$, $\delta_{7,10}-\delta_{5,10}$, $\delta_{28,15}-\delta_{6,15}$,  and $\delta_{20,20}-\delta_{10,20}$ in Figures \ref{theta}. We can observe that 45 iterations are enough for  convergence. The average running time of the algorithm for different initial conditions is $19$ seconds for $200$ random initial conditions.

 We use MOSEK solver to solve the ISO's centralized problem~(\ref{ISO_problem}). The solution is the same as the decentralized approach, but the  running time is 35 seconds. To further elaborate the comparison, we provide  the average running time of  Algorithm 1 and the centralized approach for six test systems \cite{ieee1} in  Fig. \ref{run}, we also use the approach in \cite{pedramshahab}  that applies semidefinite programming (SDP) to obtain the global optimal  point in a grid using AC OPF analysis and the CVaR as the risk measure. In addition to comparing the algorithm running time, the global optimality of the solution to the AC OPF enables us to quantify the approximation in using the DC OPF in our decentralized algorithm.  The main difference is that the AC OPF
 includes the network losses, while the DC OPF  doesn not consider the network losses. Furthermore, the SDP approach in \cite{pedramshahab} returns the global optimal solution to the AC OPF.  The calculation results show that the values of the ISO's objective using Algorithm 1 with DC OPF are lower by $3\%$ (in Polish 2383wp) to $8\%$ (in IEEE 300-bus system) than the centralized method with AC OPF and SDP approach due to the inclusion of losses and optimality of SDP method. Whereas, the computation time is much lower in Algorithm 1.

\begin{table}[t]  \vspace*{-1mm}\centering\footnotesize  \caption{The optimal value and Average CPU Time for the Deterministic Multi-stage  Algorithm and  Our Proposed  Algorithm.}\label{compare}
    \begin{tabular}{ |c|c|c|c|c|}\cline{2-5}
\multicolumn{1}{c}{}&\multicolumn{2}{|c|}{Algorithm 1}&\multicolumn{2}{c|}{\specialcell{Centralized algorithm\\with AC OPF}}\rule{0pt}{4ex} \\ [0.5ex]\hline
        $\!\!\!\!\text{\!\!\!\!\!Test system}\!\!\!\!$  &$\!\!\!\!\!f^{\text{ISO}}\,(\$)\!\!\!\!\!\!$& $\!\!\!\text{CPU time\,(s)\!\!\!} $ &$\!\!f^{\text{ISO}} \,(\$)\!\!\!\!\!\!\!\!$ & $\!\!\!\text{CPU time\,(s)}\!\!\!\!\!$ \rule{0pt}{2.8ex}  \\[0.5ex]\hline 
$\!\!\text{IEEE 14-bus}\!\!\!\!$& $\!\!\!207,\!780.8\!\!$ &$\!10$ & $\!\!\!217,\!991.1\!\!\!$ &$\!24$ \rule{0pt}{3ex} \\ [0.5ex] \hline        
        $\!\!\!\!\text{IEEE 30-bus}\!\!$& $\!\!\!330,\!760.2\!\!$ &$\!16$ & $\!\!\!347,\!980.8\!\!\!$ &$\!63$ \rule{0pt}{3ex} \\ [0.5ex] \hline
        $\!\!\!\!\text{IEEE 118-bus}\!\!$& $\!\!\!3,\!290,\!813.4\!\!$ &$\!25$ &  $\!\!\!3,\!490,\!815.8\!\!\!$ &$\!153$ \rule{0pt}{3ex} \\ [0.5ex] \hline
      $\!\!\!\!\text{IEEE 300-bus}\!\!$& $\!\!\!18,\!152,\!944.3\!\!$ &$\!42$ & $\!\!\!19,\!845,\!152.8\!\!\!$ &$\!218$ \rule{0pt}{3ex} \\ [0.5ex] \hline
      $\!\!\!\text{Polish 2383wp}\!\!$& $\!\!\!43,\!263,\!027.9\!\!$ &$\!136$ & $\!\!\!44,\!164,\!101.3\!\!\!$ &$\!1,\!312$ \rule{0pt}{3ex} \\ [0.5ex] \hline
      $\!\!\!\text{Polish 3012wp}\!\!$& $\!\!\!64,\!121,\!518.3\!\!\!$ &$\!182$ & $\!\!\!71,\!426,\!241.2\!\!\!$ &$\!1,\!636$ \rule{0pt}{3ex}\\ [0.5ex]  \hline
    \end{tabular}
    \end{table}

  \vspace{-3mm}
\section{Conclusion} \label{s6}

 In this paper, we proposed a decentralized algorithm for energy trading among load aggregators and generators in a power grid. In our model, the ISO sends control signals to the entities to motivate them towards  optimizing their objectives independently, while meeting the physical constraints of the power network. We  introduced the concept of CVaR to limit the likelihood of renewable generation shortage in the day-ahead market. To evaluate the performance of the proposed decentralized algorithm, we used an IEEE 30-bus test system connected to some renewable generators. Results confirmed that the algorithm converges in 45 iterations. We also evaluated the  price responsive load  profiles and generation values for various  discomfort costs of the load aggregators, and showed that the proposed decentralized algorithm can benefit the load aggregators by reducing their cost by $18\%$, and the generators by reducing the PAR by $15\%$ and increasing their profit by $17.1\%$. Our algorithm benefit the ISO by maintaining the privacy issues and a lower computational time compared to the centralized approach. When compared with  a centralized method with AC power flow equations, our approach has a lower running time at the cost of $3\%$ to $8\%$ approximation error due to using the DC power flow equations.

\bibliography{mybibfile,ref,mybibfile2,PESRef}
\vspace{-6mm}
 \bibliographystyle{IEEEtran}
\end{document}